\def\nk{n_{\rm b}}

\def\rfr#1{eq. (\ref{#1})}

\def\derp#1#2{\rp{\partial{#1}}{\partial{#2}}}
\def\dert#1#2{\frac{{{d}}{#1}}{{{d}}{#2}}}

\def\virg#1{``#1''}

\def\eqi{\begin{equation}}
\def\eqf{\end{equation}}
\def\eqia{\begin{eqnarray}}
\def\eqfa{\end{eqnarray}}
\def\Om{\mathit{\Omega}}
\def\rp#1#2{{#1\over#2}}
\def\lb#1{\label{#1}}

\def\bds#1{\boldsymbol{#1}}

\def\cI{\cos I}

\def\ton#1{\left(#1\right)}
\def\qua#1{\left[#1\right]}
\def\grf#1{\left\{#1\right\}}
\def\ang#1{\left\langle #1\right\rangle}

\documentclass[Galaxies,article,accept,oneauthor,dvipdfm,12pt,a4paper]{mdpi} 

\lastpage{x}
\doinum{10.3390/------}
\pubvolume{xx}
\pubyear{2013}
\history{Received: xx / Accepted: xx / Published: xx}
\usepackage{amsmath,amssymb,amsthm,amscd,latexsym}
\usepackage{hyperref}
\usepackage[toc,title,titletoc]{appendix}
\usepackage{graphicx,epsfig}
\RequirePackage{color}

\Title{Two-body orbit expansion  due to time-dependent relative acceleration rate of the cosmological scale factor}

\Author{Lorenzo Iorio $^{1,}$*}

\address{%
$^{1}$ Italian Ministry of Education, University and Research (M.I.U.R.)-Education, Fellow of the Royal Astronomical Society (F.R.A.S.), Viale Unit\`{a} di Italia 68, 70125, Bari (BA), Italy. Tel. +39 3292399167}

\corres{lorenzo.iorio@libero.it}

\abstract{
By phenomenologically assuming a slow temporal variation of the percent acceleration rate $\ddot S S^{-1}$ of the cosmic scale factor $S(t)$, it is shown that the orbit of a local binary undergoes a secular expansion.  To first order in the power expansion of $\ddot S S^{-1}$ around the present epoch $t_0$, a non-vanishing shift per orbit $\ang{\Delta r}$ of the two-body relative distance $r$ occurs for eccentric trajectories. A general relativistic expression, which turns out to be cubic in the Hubble parameter $H_0$ at the present epoch, is explicitly calculated for it in the case of matter-dominated epochs with Dark Energy.  For a highly eccentric  Oort comet orbit with  period $P_{\rm b}\approx 31$ Myr, the general relativistic distance shift per orbit turns out to be of the order of $\ang{\Delta r}\approx 70$ km. For the Large Magellanic Cloud, assumed on a bound elliptic orbit around the Milky Way, the shift per orbit is of the order of $\ang{\Delta r}\approx 2-4$ pc. Our result has a  general validity since it holds in any cosmological model admitting the Hubble law and a slowly varying $\ddot S S^{-1}(t)$. More  generally, it is valid  for an arbitrary  Hooke-like extra-acceleration whose \virg{elastic} parameter $\mathcal{K}$ is slowly time-dependent, irrespectively of the physical mechanism which may lead to it. The coefficient $\mathcal{K}_1$ of the first-order term of the power expansion of $\mathcal{K}(t)$ can be preliminarily constrained in a model-independent way down to a $\mathcal{K}_1\lesssim 2\times 10^{-13}$  year$^{-3}$ level from latest Solar System's planetary observations. The radial velocities of the double
lined spectroscopic binary $\alpha$ Cen AB yield $\mathcal{K}_1\lesssim  10^{-8}$ year$^{-3}$.
}



\begin{document}

%

\section{Introduction}

In standard cosmology, the expansion of the Universe affects the dynamics of a localized gravitationally bound two-body system at the Newtonian level with an extra-acceleration ${\bds A}_{\rm cos}$ of the order $\mathcal{O}\ton{H^2}$ in the Hubble parameter $H(t)$. For recent reviews, see e.g. \citep{2010RvMP...82..169C,2013arXiv1306.0374G} and references therein.
Let us recall that, from the Hubble law
\eqi \dot{\bds r} = H(t)\bds r, \eqf
it follows
\eqi\ddot {\bds r} = \dot H\bds r +  H\dot{\bds r} = \ton{\dot H +  H^2}{\bds r},\lb{hubacc}\eqf
where $r$ is the relative distance of the binary considered.
Since the Hubble parameter $H(t)$ is defined at any time as
\eqi H(t)\doteq\rp{\dot S}{S},\eqf where $S(t)$ is the cosmological scale factor,
\rfr{hubacc} reduces to
\eqi{\bds A}_{\rm cos} = \ton{\rp{\ddot S}{S}}\bds r = -q H^2{\bds r},\lb{hubAcc}\eqf
where the dimensionless deceleration parameter is usually defined as
\eqi q\doteq -\rp{1}{H^2}\ton{\rp{\ddot S}{S}}.\eqf
Let us rewrite the Hooke-like acceleration of \rfr{hubAcc} as \eqi{\bds A}_{\rm cos} = \mathcal{K}\bds r,\lb{accel}\eqf
with
\eqi \mathcal{K}\doteq \rp{\ddot{S}}{S}.\eqf
As far as the impact of \rfr{hubAcc} on the dynamics of a local binary are concerned, the constancy of the \virg{elastic} parameter $\mathcal{K}$ has always been assumed so far in the literature.
The resulting orbital effects of \rfr{accel}, calculated with a variety of approaches (see, e.g., \citep{1998ApJ...503...61C,2007CQGra..24.5031M,2007PhRvD..75f4031S,2007PhRvD..75f4011A,2012MNRAS.422.2931N}), do not imply an expansion of the binary's orbit itself, which only undergoes a secular  precession $\ang{\dot\omega}$, where the angle $\omega$ is the argument of periapsis. A velocity-dependent acceleration of the order $\mathcal{O}\ton{H}$, causing an orbit expansion \citep{2013MNRAS.429..915I}, occurs at the first post-Newtonian (1PN) level \citep{2012PhRvD..86f4004K}.

Actually, the scale factor's relative acceleration rate $\ddot S S^{-1}$ is, in general, time-dependent, as implied by several cosmological scenarios \citep{2002ApJ...569...18T, 2006Ap&SS.306...11P, 2007Ap&SS.311..413P, 2011PhR...509..167C, 2012IJTP...51..612A, 2012PhR...513....1C}.
In this paper, we want to explore the consequences, at the Newtonian level, of a slow temporal variation of $\mathcal{K}$ on the orbital dynamics of a localized gravitationally bound restricted two-body system. First, we will calculate them within standard general relativistic cosmology for a matter-dominated era with Dark Energy in a flat Universe. Then, we will also use latest data from Solar System planetary dynamics to phenomenologically infer preliminary constraints on the parameter of the time-dependent acceleration.
\section{General Relativistic Orbit Expansion  in the  Era Dominated by Non-Relativistic Matter and Dark Energy}

At present, the simplest cosmological model providing a reasonably good match to many different kinds of  observations is the so-called
$\Lambda$CDM model; in addition to the standard forms of baryonic matter and radiation, it also implies the existence of the Dark Energy (DE), accounted for by a cosmological constant $\Lambda$, and of the non-baryonic cold Dark Matter (DM). It assumes  general relativity as the correct theory of the gravitational interaction at cosmological scales.
The first Friedmann equation for a Friedmann-Lema\^{\i}tre-Roberston-Walker (FLRW) spacetime metric describing a homogenous and isotropic non-empty Universe endowed with a cosmological constant $\Lambda$ is \citep{Pad010}
\eqi \ton{\rp{\dot S}{S}}^2 + \rp{k}{S^2}= H_0^2\qua{\Omega_{\rm R}\ton{\rp{S_0}{S}}^4 + \Omega_{\rm NR}\ton{\rp{S_0}{S}}^3+\Omega_{\Lambda}},\lb{lunga}\eqf
where $k$ characterizes the curvature of the spatial hypersurfaces, $S_0$ is the present-day value of  the expansion scale factor, and the dimensionless energy densities $\Omega_{i}, i={\rm R, NR},\Lambda$, normalized to the critical energy density \eqi\varepsilon_{\rm c} = \rp{3c^2 H_0^2}{8\pi G},\eqf where $G$ is the Newtonian constant of gravitation and $c$ is the speed of light in vacuum,  refer to their values at $S = S_0$. Based on the equation of state relating the pressure $p$ to the energy density $\varepsilon$ of each component, $\Omega_{\rm R}$ refers to the relativistic matter characterized by $p_{\rm R}=(1/3)\varepsilon_{\rm R}$, $\Omega_{\rm NR}$ is the sum of  the normalized energy densities of the ordinary baryonic matter and of the non-baryonic dark matter, both non-relativistic, while $\Omega_{\Lambda}$ accounts for the dark energy modeled by the cosmological constant $\Lambda$ in such a way that $p_{\Lambda} = -\varepsilon_{\Lambda}$.
By keeping only $\Omega_{\rm NR}$ and $\Omega_{\Lambda}$ in \rfr{lunga}, it is possible to integrate it, with $k=0$, to determine $S(t)$ for epochs characterized by the accelerated expansion of the Universe. The result is \citep{Pad010}
\eqi
\rp{S(t)}{S_0} = \ton{\rp{\Omega_{\rm NR}}{\Omega_{\Lambda}}}^{1/3}\sinh^{2/3}\ton{\rp{3}{2}\sqrt{\Omega_{\Lambda}}H_0 t}.
\lb{scale}\eqf
As a check of the applicability of \rfr{scale} to the present epoch $t_0$, let us calculate \rfr{scale} by using just the current values for the parameters entering it; they are
\citep{2013arXiv1303.5076P}
\begin{align}
t_0 \lb{prima} &= \ton{13.813 \pm 0.058}\ {\rm Gyr}, \\ \nonumber \\
\Omega_{\Lambda} & = 0.686\pm 0.020, \\ \nonumber \\
H_0 \lb{ultima} & = (6.89 \pm 0.14)\times 10^{-11} \ {\rm year^{-1}}.
\end{align}
As a result, the right hand side of \rfr{scale} turns out to be equal to $0.99\pm 0.02$, which is compatible with the expected value of 1 for the left hand side of \rfr{scale} evaluated at $t=t_0$.

From \rfr{scale} it turns out
\eqi
\rp{\ddot S}{S} = -\rp{1}{2}H_0^2\Omega_{\Lambda}\qua{-3  + \coth^{2}\ton{\rp{3}{2}\sqrt{\Omega_{\Lambda}}H_0 t} },\lb{Kappa}
\eqf
which explicitly shows that, in standard general relativity, $\mathcal{K}$ is naturally time-dependent for the eons considered.
Let us, now, expand $\mathcal{K}(t)$ in powers of $t$ around $t_0$.
To the first order in $\Delta t \doteq t - t_0$, one has
\begin{align}
\mathcal{K}(t) \nonumber & \simeq \mathcal{K}_0+\mathcal{K}_1\Delta t = -\rp{1}{2}H_0^2\Omega_{\Lambda}\qua{-3  + \coth^{2}\ton{\rp{3}{2}\sqrt{\Omega_{\Lambda}}H_0 t_0} } + \\ \nonumber \\
& + \rp{3}{2} H_0^3\Omega_{\Lambda}^{3/2}\coth\ton{\rp{3}{2}\sqrt{\Omega_{\Lambda}}H_0 t_0} {\rm csch}^2\ton{\rp{3}{2}\sqrt{\Omega_{\Lambda}}H_0 t_0}\Delta t.\lb{approx}
\end{align}
Thus, we pose
\begin{align}
{\bds A}_0 \lb{accel0} & = \mathcal{K}_0\bds r, \\ \nonumber \\
{\bds A}_1 \lb{accel1} & = \mathcal{K}_1 \Delta t ~ \bds r.
\end{align}
According to \rfr{prima}-\rfr{ultima}, \rfr{approx} yields
\begin{align}
\mathcal{K}_0 &= 2.5\times 10^{-21}\ {\rm year^{-2}}, \\ \nonumber \\
\mathcal{K}_1 \lb{piccolo} &= 1.5\times 10^{-31}\ {\rm year^{-3}}.
\end{align}

Let us, now, consider a localized gravitationally bound two-body system over time intervals $|\Delta t|$ small enough to look at the extra-acceleration of \rfr{accel1}, evaluated with \rfr{approx}, as a small correction to the standard Newtonian monopole. Given the figures in \rfr{prima}-\rfr{ultima}, such a condition is satisfied for a variety of binaries and timescales; for example, the Newtonian acceleration of an Oort comet at $10^5$ au from the Sun is several orders of magnitude larger that \rfr{accel1} evaluated for $\Delta t = 5$ Gyr. Thus, it is possible to work out perturbatively the long-term orbital effects induced by \rfr{accel1}.
%
%
%
%
%
%
%
%
%
%

The shift per orbit of the distance $r$ induced by a generic perturbing acceleration  can be calculated as
\eqi \ang{\Delta r} \lb{Deltaerre}= \int_0^{P_{\rm b}}dr = \int_0^{P_{\rm b}}\dot r dt = \int_0^{P_{\rm b}} \ton{\derp r E \dert E {\mathcal{M}}\dert{\mathcal{M}} t + \derp r a \dert a t+ \derp r e \dert e t} dt,\eqf where $E,\mathcal{M},a,e$ are the eccentric anomaly,  the mean anomaly,  the semimajor axis, and  the eccentricity, respectively. Moreover, $P_{\rm b}=2\pi\sqrt{a^3 G^{-1} M^{-1}}$ is the unperturbed orbital period.
The integrand of \rfr{Deltaerre} has to be evaluated onto the Keplerian ellipse assumed as unperturbed trajectory. For it, the following relations hold\footnote{For computational purposes, it turns out more convenient to use the eccentric anomaly $E$ as fast variable of integration instead of the true anomaly $f$.}
\begin{align}
r \lb{erre} &= a\ton{1 - e\cos E}, \\ \nonumber \\
dt \lb{dt} & = \ton{\rp{1-e\cos E}{\nk}}dE, \\ \nonumber \\
\Delta t \lb{Deltat} &= \rp{E - e\sin E}{\nk}, \\ \nonumber \\
\sin f & = \rp{\sqrt{1-e^2}\sin E}{1 - e\cos E}, \\ \nonumber \\
\cos f & = \rp{\cos E - e}{1- e \cos E}, \\ \nonumber \\
\dert E{\mathcal{M}} \lb{dEdM} & = \rp{1}{1-e\cos E},
\end{align}
where $\nk \doteq 2\pi P^{-1}_{\rm b}$ is the unperturbed Keplerian mean motion.
In \rfr{Deltaerre}, \rfr{erre}-\rfr{dt} are used to compute the partial derivatives of $r$ and to express $dt$ in terms of $dE$, respectively,  while \rfr{dEdM} is for $dE/d{\mathcal{M}}$. Concerning the instantaneous values of $\dot a,\dot e, \dot{\mathcal{M}}$ entering \rfr{Deltaerre}, they are to be taken from the right-hand-sides of the standard Gauss equations \citep{2011rcms.book.....K} for the variation of those orbital elements.
They are \citep{2011rcms.book.....K}
\begin{align}
\dert a t \lb{dadt} & = \rp{2}{\nk\sqrt{1-e^2}}\qua{e A_r\sin f  + \ton{\rp{p}{r}}A_{\tau}}, \\ \nonumber \\
\dert e t \lb{dedt} & = \rp{\sqrt{1-e^2}}{\nk a}\grf{A_r\sin f + A_{\tau}\qua{\cos f + \rp{1}{e}\ton{1 - \rp{r}{a}} } }, \\ \nonumber \\
\dert\omega t \lb{dpdt} & = \rp{\sqrt{1-e^2}}{\nk a e}\qua{-A_r \cos f + A_{\tau}\ton{1 + \rp{r}{p}} \sin f} - \cI\dert\Om t, \\ \nonumber \\
\dert{\mathcal{M}} t \lb{dMdt} & = \nk -\rp{2}{\nk a}A_r\ton{\rp{r}{a}} -\sqrt{1-e^2}\ton{\dert\omega t  + \cI\dert\Om t},
\end{align}
where $p=a(1-e^2)$ is the semilatus rectum,   and $A_r,A_{\tau}$ are the radial and transverse components of the perturbing acceleration, respectively.
In general, the Gauss equations are applicable to whatsoever perturbing acceleration, irrespectively of its physical origin.
In our case, \rfr{accel1} is entirely radial, so that $A_{\tau}=A_{\nu}=0,$ where $A_{\nu}$ is the out-of-plane component of the perturbing acceleration. This also implies that neither the inclination $I$ nor the longitude of the ascending node $\Om$ are changed since their rates are proportional to $A_{\nu}$ \citep{2011rcms.book.....K}.
It is intended that the right-hand-sides of \rfr{dadt}-\rfr{dMdt}, when inserted in \rfr{Deltaerre}, have to be evaluated onto the unperturbed Keplerian ellipse.

It turns out that \rfr{approx} and \rfr{accel1}, applied to \rfr{Deltaerre}, yield\footnote{An analogous calculation for $\mathcal{K}_0$ in \rfr{accel0} returns $\ang{\Delta r}=0$.}
\eqi\ang{\Delta r} = \rp{\pi a e\ton{1 - e^2}\ton{16 + 9e} H_0^3\Omega_{\Lambda}^{3/2}\coth\ton{\rp{3}{2}\sqrt{\Omega_{\Lambda}}H_0 t_0} {\rm csch}^2\ton{\rp{3}{2}\sqrt{\Omega_{\Lambda}}H_0 t_0}}{8\nk^3}.\lb{derre}\eqf
Note that \rfr{derre} is positive, i.e. the distance increases. Moreover, \rfr{derre} vanishes for circular orbits. For an Oort comet orbiting the Sun along a highly eccentric ($e=0.98$) orbit in some $P_{\rm b} \approx 31$ Myr, \rfr{derre} yields a shift per orbit of about $\ang{\Delta r} \approx 70$ km.
It should be pointed out that such an effect of cosmological origin would be subdominant with respect to the consequences of the Galactic tide \citep{1986Icar...65...13H}, as suggested\footnote{I am grateful to John D. Barrow for having pointed it to me.} by a naive order-of-magnitude calculation. Indeed, from \rfr{accel1} and \rfr{piccolo}, it turns out that $A_1\approx 10^{-23}$ m s$^{-2}$ for the aforementioned Oort comet. On the other hand, the magnitude of its Galactic tidal acceleration \citep{2005CeMDA..93..229F} is of the order of \citep{2001AJ....121.2253L, 2012MNRAS.419.2226I} $A_{\rm tid}\approx 10^{-14}$ m s$^{-2}$.
By assuming a bound orbit for the Large Magellanic Cloud (LMC) about the Milky Way (MW) with $P_{\rm b}\approx 3-4$ Gyr and $e\approx 0.6-0.7$ \citep{2013ApJ...764..161K}, the corresponding  shift per orbit is roughly of the order of $\ang{\Delta r}\approx 2-4$ pc.

In principle, the calculation yielding \rfr{derre} retains its validity for a generic Hook-type extra-acceleration of the form of \rfr{accel} whose \virg{elastic} parameter $\mathcal{K}$ is slowly time-dependent, independently of its physical origin. In a cosmological context, different models may well lead to various analytical forms of the relative acceleration rate of the scale parameter with respect to \rfr{Kappa}.
Thus, \rfr{derre} confirms that, in an  expanding Universe, localized gravitationally bound systems slowly vary their sizes provided that the time-dependence of the relative acceleration rate of the cosmic scale factor is properly taken into account.
\section{Phenomenological Bounds from Astronomical Observations}

Here, we will adopt a phenomenological approach to preliminary infer upper bounds on the parameter $\mathcal{K}_1$ of the time-dependent term of the Hooke-type acceleration  in a model-independent way. More specifically, we will start from \rfr{accel1} without specifying any theoretical prediction for $\mathcal{K}_1$ which, thus, will be assumed as a free parameter to be constrained from observations.

Among the long-term orbital effects caused by \rfr{accel1}, there is also a secular precession of the pericenter $\varpi$. From \rfr{dpdt}, it turns out to be
\eqi\ang{\dot\varpi} = \rp{3\pi {\mathcal{K}_1}\sqrt{1-e^2}}{2\nk^2}.\lb{dodt}\eqf
This fact allows us to use the latest constraints $\Delta\dot\varpi$ on the anomalous perihelion precessions of some planets of the Solar System \citep{2011CeMDA.111..363F, 2013AstL...39..141P, 2013MNRAS.432.3431P} to infer preliminary upper bounds on $\mathcal{K}_1$. Strictly speaking, our figures are not constraints in the sense that, in \citep{2011CeMDA.111..363F, 2013AstL...39..141P, 2013MNRAS.432.3431P}, only standard general relativity was included in the dynamical models fit to the planetary observations;  $\mathcal{K}_1$ was not estimated in a least-square sense as a solve-for parameter in the global solutions of \citep{2011CeMDA.111..363F, 2013AstL...39..141P, 2013MNRAS.432.3431P} giving $\Delta\dot\varpi$. As such, our bounds should rather be seen as an indication of acceptable values, given the state-of-the-art in the field of planetary ephemerides.
From $\nk^{-2}$ in \rfr{dodt}, it turns out that wide orbits are good candidates for our purposes. Thus, we will use the most recent determinations for the perihelion of Saturn, whose anomalous precession has been constrained down to\footnote{In \rfr{pit}, mas cty$^{-1}$ stands for milliarcseconds per century.}  \citep{2013AstL...39..141P, 2013MNRAS.432.3431P} \eqi \Delta\dot\varpi \leq 0.47\ {\rm mas\ cty^{-1}}\lb{pit}\eqf with the EPM2011 ephemerides \citep{2013SoSyR..47..386P}.

By comparing \rfr{dodt}, calculated for Saturn, with \rfr{pit}, yields
\eqi \mathcal{K}_1\lesssim 2\times 10^{-13}\ {\rm year^{-3}}. \lb{saturn}\eqf

In principle, another local astronomical laboratory which could be used to infer bounds on $\mathcal{K}_1$ is the double lined spectroscopic binary system $\alpha$ Cen AB \citep{2002A&A...386..280P}, which has an orbital period as long as $P_{\rm b} = 79.91$ year and an eccentricity of $e = 0.5179$. In this case, the observable quantities are  the radial velocities of both A and B, known with an accuracy of about $\approx 4-7$ m s$^{-1}$ \citep{2002A&A...386..280P}. By analytically calculating the shift per orbit of the radial velocity due to \rfr{accel1}, it turns out
\eqi \mathcal{K}_1 \lesssim 10^{-8}\ {\rm  year^{-3}},\eqf
which is not competitive with \rfr{saturn}.
\section{Summary and Conclusions}

In any cosmological model admitting the Hubble law, a Hooke-type acceleration naturally arises for a localized gravitationally bound two-body system. Its \virg{elastic} coefficient $\mathcal{K}$ is the relative acceleration rate $\ddot S S^{-1}$ of the cosmic scale factor $S(t)$. If a slow temporal variation is assumed for it, a power expansion to first order around the present epoch $t_0$ yields, among other things, a net increase $\ang{\Delta r}$ of the orbital size.

In standard general relativity with a cosmological constant, it is cubic in the ratio of the Hubble parameter
 $H_0$ at the present epoch to the binary's orbital frequency $\nk$. By using its general relativistic expression for epochs dominated by non-relativistic matter and Dark Energy, such an expansion, occurring for non-circular orbits, is of the order of $\ang{\Delta r} \approx 2-4$ pc  for the Large Magellanic Cloud, assumed orbiting the Milky Way along an elliptic path with a period $P_{\rm b}$ of a few Gyr. Quite smaller effects arise for Solar System's objects such as, e.g., an Oort comet with a period of $P_{\rm b}\approx 31$ Myr: its trajectory expands by just $\ang{\Delta r}\approx 70$ km per orbit.

 Our calculation is quite general since it is valid, in principle, also for other models showing a different functional temporal dependence of the relative acceleration rate of the cosmological scale factor. Moreover, it holds also for any putative Hooke-like non-standard acceleration with time-dependent elastic parameter $\mathcal{K}(t)$, irrespectively of its physical origin. We phenomenologically infer upper bounds on the coefficient $\mathcal{K}_1$ of the first-order term of its power expansion from latest Solar System planetary observations. The existing constraints on the anomalous perihelion precession $\Delta\dot\varpi$ of Saturn allows to obtain $\mathcal{K}_1\lesssim 2\times 10^{-13}$ year$^{-3}$. The radial velocities of the double lined spectroscopic binary  $\alpha$ Cen AB, with a period of $P_{\rm b} = 79.91$ year, yield $\mathcal{K}_1\lesssim 10^{-8}$ year$^{-3}$.
%
%
%
%
%
%
%
%
%
\section*{Acknowledgments}
I thank John D. Barrow and Salvatore Capozziello for useful correspondence.

\bibliography{Hubblebib,PPN_alpha3bib,pfebib,ephemeridesbib,Youngsunbib}{}
\bibliographystyle{mdpi-arXiv}
\end{document}